# QUANTUM WALK INSPIRED JPEG COMPRESSION OF IMAGES

Abhishek Verma[1], Sahil Tomar[2], and *Sandeep Kumar[3]

[1,2,3] Central Research Laboratory, BEL-Ghaziabad, India

Corresponding Author: sann.kaushik@gmail.com

*Abstract*—This work proposes a quantum-inspired adaptive quantization framework that enhances the classical JPEG compression by introducing a learned, optimized Q-table derived using a Quantum Walk Inspired Optimization (QWIO) search strategy. The optimizer searches a continuous parameter space of frequency-band scaling factors under a unified rate–distortion objective that jointly considers reconstruction fidelity and compression efficiency. The proposed framework is evaluated on MNIST, CIFAR-10, and ImageNet subsets, using Peak Signal-to-Noise Ratio (PSNR), Structural Similarity Index (SSIM), Bits-Per-Pixel (BPP), and error-heatmap visual analysis as evaluation metrics. Experimental results show average gains ranging from 3–6 dB PSNR, along with better structural preservation of edges, contours, and luminance transitions, without modifying decoder compatibility. The structure remains JPEG compliant and can be implemented using accessible scientific packages making it ideal for deployment and practical research use.

## I. INTRODUCTION

The rapid growth of digital imaging across domains such as remote sensing, biomedical archives, surveillance networks, multimedia platforms, and large-scale visual datasets has made image compression a critical component of modern data storage and transmission systems. Compression enables efficient transmission over constrained bandwidth channels and reduces storage cost while maintaining visual quality acceptable to human perception [1]. Image compression techniques can be broadly categorized into lossless and lossy methods. Lossless compression techniques, such as PNG, GIF, or JPEG-LS, preserve exact pixel values, but their compression ratios are typically limited to 2:1 or 3:1. On the other hand, lossy compression techniques like JPEG, JPEG2000, and HEVC-based image coding etc. achieve far higher compression ratios by removing redundancies and perceptually insignificant details [2].

Lossy compression relies on the psychovisual redundancy present in images: the human eye is less sensitive to high-frequency information such as fine textures or minor color fluctuations. By selectively discarding or coarsely quantizing such information, lossy compression reduces file size substantially while preserving the overall perceptual quality. The successful reconstruction of such files can play a decisive role in the results of crucial investigations [3].

Out of all the compression standards developed in the last 3 decades, JPEG [4] is the most popular compression technique, because of its low computational complexity, hardware compatibility, and long-standing industry support. Its encoding pipeline, based on block-wise Discrete Cosine Transform (DCT) [5], luminance–chrominance separation, and quantization-driven redundancy reduction, continues to provide an effective balance between file size reduction and perceptual image quality for a wide spectrum of visual data.

However, despite its enduring success, the classical JPEG framework still relies on a manually designed, static quantization matrix that does not adapt to differences in image structure, texture distribution, or dataset-specific frequency characteristics. This lack of adaptivity leads to sub-optimal rate–distortion (RD) performance, poor customization and visible degradation when operating at higher compression levels.

The quantization matrix is crucial in defining how coefficients across different frequency bands are preserved, suppressed, or discarded during compression. The standard JPEG luminance quantization table was derived empirically through early perceptual experiments conducted on limited natural image samples, and thus reflects a fixed compromise between compression strength and visual reconstruction quality [6]. However, images exhibit a wide range of structural and spectral characteristics; datasets consisting of low-texture medical images, handwritten digits, or structured natural scenes respond very differently to quantization. In some datasets, high-frequency components may correspond primarily to noise, while in others they represent subtle transitions in shading, contours, textures, and fine structural details that are essential for visual interpretation. This observation agrees with recent comprehensive analyses in image restoration, which emphasize that frequency components play different roles depending on image content and application context, further motivating the need for adaptive and content-aware compression strategies [7].

Over the years, multiple attempts have been made to redesign or tune JPEG quantization matrices using heuristic search, evolutionary computing, RD modeling, or dataset-conditioned adaptation etc. The work proposed in [8] has shown improvement in Peak Signal-to-Noise Ratio (PSNR) but remain limited by high computational complexity. Evolutionary multi-objective optimization method proposed in [9] achieved strong RD performance and flexible bit rate adaptation, yet required pre-defined statistical lookup tables and multistage population management, which increased algorithmic complexity. Similarly, genetic-algorithm based optimization for iris image compression in [10] produced highly specialized quantization table but optimization was application specific and not applicable on general purpose image dataset. More recent literature has explored quantum-inspired search mechanisms as a means of improving exploration efficiency in complex optimization landscapes, but most such efforts have been theoretical in nature, with limited alignment to practical, standard-compliant compression workflows [11].

Despite extensive advancements in image compression techniques, the JPEG standard still lacks an effective mechanism for automatically adapting its quantization behavior to different image structures, dataset characteristics, and RD requirements. Moreover, it lacks an effective mechanism to automatically learn quantization behaviour that adapts to image content, dataset characteristics, and RD objectives, while still preserving compatibility with existing decoders [4]. Existing optimization approaches either operate in discrete search spaces prone to premature convergence, require heavy computational resources, or rely on learning-based architectures that are difficult to integrate into legacy JPEG pipelines. On the other hand, learning-based compression models typically require complex architectures and are difficult to integrate into legacy JPEG pipelines. As a result, there remains a practical challenge in designing a quantization learning framework that is continuous, stable, computationally efficient, and fully compatible with standard JPEG decoding [6]. This study addresses this problem by formulating a quantum-walk-inspired optimization strategy that enables adaptive, interpretable, and dataset-aware quantization learning within the classical JPEG compression framework.

## II. Literature Review

Image compression is an essential process in modern digital imaging systems, serving as the foundation for efficient image storage, transmission, and processing etc. Transform-based coding has been the root of modern image compression, where visual information is mapped into a frequency domain representation such that spatial redundancy and perceptual irrelevance are eventually exploited. The work given in [5] proposed the use of DCT to provide a compact representation of image signals with high energy concentration that subsequently allowed the development of the JPEG standard. The baseline JPEG architecture, formalized by the Joint Photographic Experts Group [4], applies block-wise DCT, luminance–chrominance separation, coefficient quantization, and entropy coding in its pipeline to achieve significant data reduction while preserving perceptual acceptability. The work proposed in [12] on wavelet theory further gave inspiration for multilayer frequency representations; hence, improved compression models, such as the JPEG 2000 [13], demonstrated superior RD performance and scalability by using discrete wavelet transforms. Despite these advances, the classical JPEG standard retained dominance due to its computational simplicity and widespread hardware support, whereas newer codecs required greater processing resources and lacked backward compatibility.

A major limitation of the JPEG framework lies in the fact that it uses a fixed quantization matrix obtained originally from heuristic perceptual tuning instead of data-driven optimization. Early research exploring adaptive or image-conditioned quantization, such as that of [14], demonstrated that adjusting quantization parameters based on image content could lead to increased quality in compression, at the cost of heavier computational load and parameter transmission overhead. Other representative optimization methodologies considered the utilization of heuristic and stochastic search methods, including simulated annealing [15] and particle swarm optimization [16], for adjusting quantization coefficients with the aim of reducing reconstruction error or RD cost. While these methods introduced meaningful improvement, they often faced slow convergence, parameter sensitivity, and difficulty operating effectively within high-dimensional discrete search spaces.

Parallel progress in large-scale datasets and learning frameworks has encouraged compression research to move toward statistical and neural models, where convolutional networks and deep autoencoders are explored as substitutes for traditional transform coding to learn latent features and perform end-to-end RD optimization [17]. Although these learned codecs may outperform JPEG in controlled scenarios, their broader use in resource-constrained or legacy environments is limited due to extensive training demands, compatibility issues, and heavy computation. Complementary studies on double JPEG compression [18] and coefficient behavior further indicate that quantization artifacts are strongly dependent on underlying parameter patterns, motivating quantization-aware optimization research.

More recently, research interest has shifted toward quantum and quantum-inspired computation to address challenging optimization issues for image compression. These studies focus on the conceptual mapping of JPEG operations into quantum frameworks and transform domains. The study proposed in [19] introduced a framework where quantum information theory is applied to JPEG compression, enabling efficient encoding of DCT coefficients while preserving perceptual quality. Similarly, a quantum-based JPEG model [20] was proposed, representing pixels as quantum states and applying quantum transformations, achieving reduced computational complexity and higher compression ratios. In the research work, Quantum image compression algorithms [11] quantum Fourier and wavelet transforms techniques to improve image storage and reconstruction were reviewed in detail. The work emphasized the potential of quantum methods for achieving compact image representations with minimal visual degradation.

The collective body of prior work emphasizes two critical observations. First, quantization remains a fundamental and highly influential component of JPEG compression, where strong potential for performance gains may be exploited by means of adaptive optimization. Second, many optimization and learning-based methods have indeed been proposed; however, most of them either operate in discrete parameter spaces that are prone to local minima, require heavy computational resources, or do not preserve compatibility with standard JPEG decoding. These limitations provide an opportunity for a lightweight, continuous, and interpretable optimization framework that can learn dataset-aware quantization structures while preserving the simplicity and interoperability of the classical JPEG architecture. The present work aims to fill these gaps by introducing a quantum-inspired optimization strategy tailored for adaptive JPEG quantization learning in a practical, standard-compliant setting.

In this work, we introduce a quantum-inspired optimization-driven compression framework, referred to as Quantum Walk Inspired Optimizer (QWIO) JPEG, in which the quantization matrix is treated as a learnable parameter set rather than a fixed empirical prior. The proposed approach replaces the static JPEG luminance quantization table with a data-driven and continuously optimized matrix obtained through a Quantum Walk Inspired Optimization strategy. The optimizer explores a continuous parameter space defined by a global scaling factor and band-wise frequency multipliers, enabling smooth and interpretable modifications to the underlying quantization structure. Candidate quantization matrices are evaluated using a RD objective that integrates reconstruction fidelity with compression efficiency, allowing the system to trade off perceptual quality and bit rate in a principled manner.

Unlike learned deep-compression frameworks or neural codecs, the proposed methodology does not alter the core JPEG pipeline, introduce new decoding dependencies, or require specialized hardware. Instead, it preserves full backward compatibility with existing JPEG decoders while adaptively redefining how frequency information is quantized based on the statistics of a particular image or dataset. The framework is further evaluated on MNIST [21], CIFAR-10 [22], Brain Tumor classification [23], CASIA-Iris [24] and ImageNet subsets to investigate its robustness across low-complexity, medium-complexity, and high-variability visual domains. The observed improvements in PSNR, Structural Similarity Index (SSIM), Bits-Per-Pixel (BPP), and spatial error-heat map characteristics collectively demonstrate that quantum-inspired optimization provides a promising, interpretable, and practically deployable pathway toward adaptive JPEG quantization learning. The novel contributions of this work are:
- A novel quantum-inspired continuous optimization framework is introduced to learn adaptive quantization matrices in a continuous search space, enabling stable convergence and improved RD performance.
- A band-based parameterization strategy that preserves frequency structure while reducing dimensionality with a RD objective function integrating mean square error (MSE) and entropy-based bitrate estimation.
- The proposed method has been extended from single image optimization toward dataset level learning and evaluated intensively on MNIST, CIFAR-10, Brain Tumor, CASIA-Iris and ImageNet datasets and proved its efficiency with improvement in PSNR and other matrices.

The rest of the paper is organized as follows. Section 3 presents the proposed QWIO-JPEG methodology, detailing the adaptive quantization formulation and the quantum-walk-inspired optimization framework. Section 4 describes the experimental setup, including dataset selection, and evaluation metrics. Section 5 discusses the quantitative and qualitative results obtained across multiple datasets. Finally, Section 6 concludes the paper.

## III. Proposed Methodology

The proposed Quantum Walk Inspired JPEG Compression framework, referred to as QWIO-JPEG, extends the classical JPEG pipeline by replacing its fixed quantization matrix with a continuously optimized, data-adaptive matrix learned through a quantum-inspired population search strategy. This methodology remains fully compliant with the baseline JPEG standard [4], meaning that the optimized quantization table can be applied without modification to the encoding or decoding architecture. Such an underpinning idea is that the quantization process represents the dominant source of distortion and bitrate reduction [5] in JPEG. Therefore, adapting the quantization coefficients to the frequency characteristics of an image or dataset can yield measurable improvements in reconstruction quality and compression efficiency.

As illustrated in Figure 1, the QWIO-JPEG framework consists of a structured optimization pipeline that integrates pre-processing and luminance-channel frequency transformation, block-wise DCT coefficient representation, parameterized adaptive quantization matrix formulation, RD-guided evaluation of candidate matrices, and a quantum-walk-inspired continuous optimization mechanism. These components together form a closed optimization loop in which candidate quantization matrices are iteratively generated, evaluated, reinforced, and refined until convergence, resulting in a final optimized quantization matrix that is subsequently reused for standard JPEG compression across the dataset. The whole methodological framework has five major parts and is explained below:

### A. Preprocessing and Luminance-Channel Frequency Transformation:

The process begins by converting the input RGB image into the YCbCr colour space, after which only the luminance ($Y$) component is used for optimization, consistent with prior findings that human perception is more sensitive to luminance distortions than chrominance deviations [4]. The luminance image is divided into non-overlapping $8 \times 8$ blocks, and the pixel intensities within each block are centered around zero, prior to transformation.

### B. Block-wise DCT and Coefficient Representation:

Each block is projected into the frequency domain using the two-dimensional DCT, which compactly concentrates energy into a small number of low-frequency coefficients as originally formulated in the study [5]. For a spatial block $B(x,y)$, the forward DCT is mathematically represented using [5] as:

$$c(u,v) = \alpha(u)\alpha(v) \sum_{x=0}^{7}\sum_{x=0}^{7} B(x,y) \cos\left(\frac{\pi(2x+1)u}{16}\right) \cos\left(\frac{\pi(2y+1)v}{16}\right) \quad (1)$$

for $0 \leq u,v \leq 7$, where $\alpha(\cdot)$ denotes the orthonormal scaling coefficient associated with the DCT basis.

### C. Quantization Matrix Formulation

The quantization stage then approximates each coefficient according to the equation :

$$Z(i,j) = round\left(\frac{C(i,j)}{Q(i,j)}\right) \quad (2)$$

where $Q(i,j)$ is the quantization value at frequency location $(i,j)$. During decompression, coefficients are calculated using [4] as:

$$\hat{C}(i,j) = Z(i,j) \cdot Q(i,j) \quad (3)$$

### D. Inverse DCT Transform

The inverse DCT define the spatial block using [5] as:

$$\hat{B}(x,y) = \sum_{x=0}^{7}\sum_{x=0}^{7} \alpha(u)\,\alpha(v)\hat{C}(u,v) \cos\left(\frac{\pi(2x+1)u}{16}\right) \cos\left(\frac{\pi(2y+1)v}{16}\right) \quad (4)$$

In this formulation, the quantization matrix governs the degree of frequency suppression, with larger entries discarding greater detail but reducing storage cost. The standard JPEG quantization table was empirically derived and is not adapted to dataset-specific statistics [4] [13], which motivate its reformulation as an optimizable parameter space.

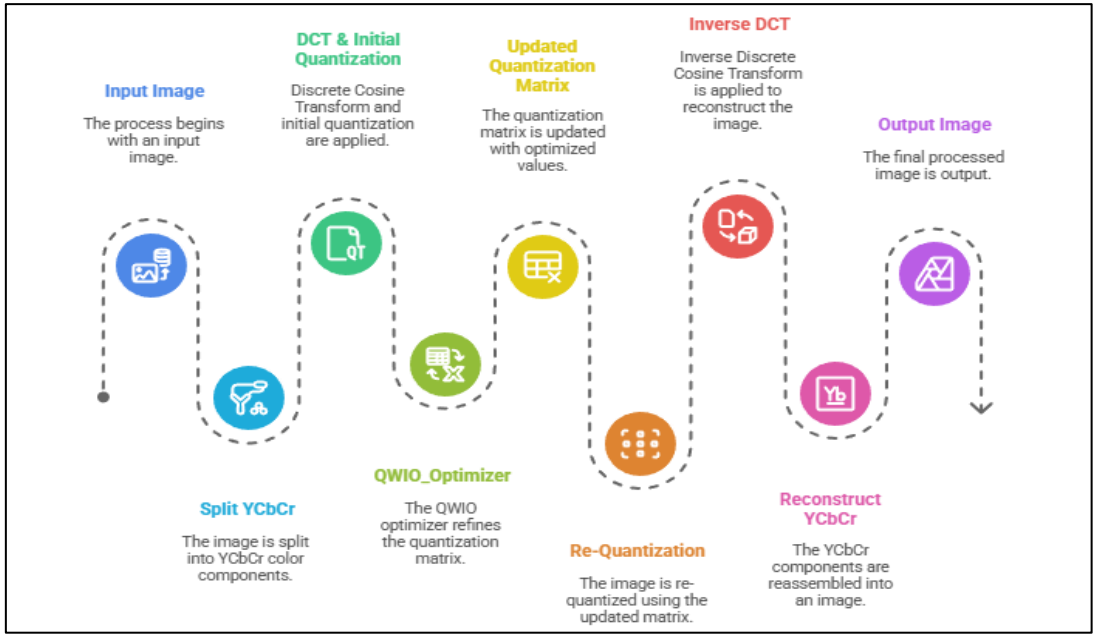

Figure 1: Information Flow Chart of the Proposed QWIO-JPEG Framework Illustrating the Integration of Quantum-Walk-Inspired Optimization With the Classical JPEG Compression Pipeline.

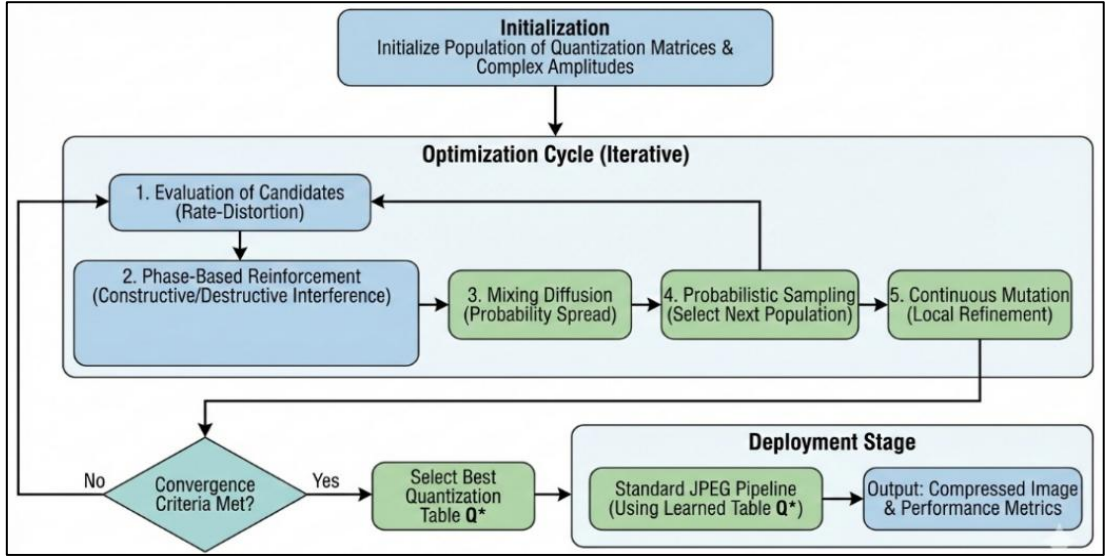

Figure 2: Flowchart of the proposed QWIO framework. The iterative process evolves candidate quantization tables using quantum walk principles: reinforcement, mixing, and sampling to learn optimal parameters for enhanced JPEG compression.

Rather than optimizing all 64 coefficients independently, which would result in an unstable and high-dimensional search problem, the proposed system employs a band-based parameterization strategy inspired by prior adaptive quantization studies [14] [25]. Coefficients are grouped according to their diagonal frequency band $k = (i + j)$, reflecting their combined horizontal–vertical spatial frequency contribution. A global scaling parameter $s$ and a set of band-wise multipliers $m_k$ modulate the baseline luminance quantization table $Q_{(i,j)}$, producing an adaptive quantizer, that can be expressed using [4] as:

$$Q(i,j) = round(s \cdot m_k \cdot Q_{(i,j)}) \qquad (5)$$

This formulation preserves the structural intent of the original JPEG quantizer while reducing the parameter space to a compact and interpretable continuous representation, enabling smooth variation of quantization strength across frequency bands instead of abrupt coefficient-wise tuning.

To evaluate each candidate quantization configuration, a RD objective is defined that jointly considers reconstruction accuracy and compression efficiency. The distortion component is measured through the mean-squared error, which is defined in [26] as:

$$MSE = \frac{1}{HW} \sum_{x=0}^{H-1} \sum_{y=0}^{W-1} \left( Y(x,y) - \hat{Y}(x,y) \right)^2 \qquad (6)$$

Where Y and $\hat{Y}$ denote the original and reconstructed luminance images. In addition to MSE, PSNR and SSIM are later reported for quality assessment, consistent with perceptual fidelity analyses in compression literature [13] [27]. The rate component is approximated via entropy-based BPP, derived from the empirical symbol distribution of the quantized coefficients following approaches used in JPEG artifact and coding behaviour analysis [18]. The entropy is computed as:

$$H = -\sum_v p(v)\log_2 p(v) \quad (7)$$

and the bit-rate estimate is given by [28] as:

$$BPP = \frac{H \cdot N}{HW} \quad (8)$$

Where, N denotes the total number of quantized coefficients. The combined optimization objective is

$$J = MSE + \lambda \cdot BPP \quad (9)$$

Where, $\lambda$ controls the balance between distortion and compression strength in alignment with classical RD formulations [14] [15].

*E. Quantum-inspired Optimization for Continuous Search over Quantization Parameters*

The principal novelty of this work lies in the formulation of a QWIO to learn the optimal quantization parameters$(s, m_k)$. The optimizer draws conceptual grounding from discrete-time quantum walks, in which a particle exists in a superposition of states whose evolution is governed by complex probability amplitudes rather than classical scalar probabilities [29]. In a quantum walk, the system state $\psi_t$ evolves according to a unitary operator, enabling constructive and destructive interference to guide probability propagation across the state space. Inspired by this behaviour, QWIO adopts an amplitude-driven population representation in which candidate solutions evolve probabilistically rather than deterministically.

As illustrated in Figure 2, the QWIO optimization process follows a structured sequence of amplitude initialization, phase-based reinforcement, mixing diffusion, probabilistic sampling, and continuous mutation, forming a closed-loop optimization cycle. This flowchart representation highlights how quantum-walk principles are mapped into a practical classical optimization algorithm that balances exploration and exploitation while maintaining stable convergence behaviour. The system state $\psi_t$ evolves according to

$$\psi_{t+1} = U\psi_t \quad (10)$$

Where, $U = S(C \otimes I)$ consist of coin operator C and a shift operator $S$. The probability of a walker occupying state x is obtained from the squared magnitude of its complex amplitude at time t is,

$$P(x,t) = |\psi(x,t)|^2 \quad (11)$$

The constructive and destructive interference redistributes probability mass in a manner that supports faster spreading than classical random walks. This interference-driven propagation motivates the amplitude-based search dynamics in QWIO.

In the proposed optimizer, each candidate quantization parameter vector is treated as an analogue of a walker state and is assigned a complex probability amplitude $a_i$. The population therefore exists in a superposition-like representation rather than a deterministic list of candidates. At iteration t, the amplitude vector is normalized such that

$$\sum_{i=1}^{N}|a_i^{(t)}|^2 = 1 \quad (12)$$

After evaluating each candidate using RD objective $J_i$, a phase-based reinforcement operator is applied. To model quantum-inspired constructive bias toward better candidates, amplitudes are updated via

$$a_i^{(t+1)} = a_i^{(t)} \cdot exp\left(-j\gamma \cdot \frac{J_i - min(J)}{max(J) - min(J) + \epsilon}\right) \quad (13)$$

Where $\gamma$ controls phase rotation strength and $\epsilon$ avoids numerical instability. Candidates with lower cost experience more constructive reinforcement, while higher-cost candidates accumulate destructive interference. This mimics the amplitude reinforcement that occurs in quantum walk constructive paths. A mixing operator is then applied to diffuse probability mass across neighboring solutions, analogous to the quantum walk shift operator:

$$\tilde{a}_i^{t+1} = \sum_{k \in N(i)} w_k a_k^{(t+1)} \quad (14)$$

Where, N(i) denotes a local neighbourhood in parameter space and $w_k$ is a Gaussian kernel weighting. This prevents population collapse and preserves exploration diversity. Sampling of the next candidate population is performed according to

$$p_i = |\tilde{a}_i^{(t+1)}|^2 \quad (15)$$

thus, mapping probability mass directly to solution selection frequency. A continuous Gaussian mutation term perturbs parameters to enable local refinement while retaining stability in the search manifold.

Across iterations, the best performing quantization matrix is preserved as $Q_{best}$ and optimization terminates upon convergence or early-stopping stability. The learned $Q_{best}$ is then saved and reused during deployment. The remaining images in the dataset are compressed using the standard JPEG pipeline, with $Q_{best}$ replacing the default luminance quantization table while chrominance channels remain unchanged to maintain compatibility. Each dataset image is encoded twice: once using the baseline JPEG table and once using the optimized learned table and reconstruction performance is evaluated using PSNR, SSIM, BPP, and pixel-wise error heat maps to visually analyze structural preservation and artifact reduction.

Through this methodology, QWIO-JPEG bridges classical transform coding with a theoretically grounded, quantum-walk-inspired amplitude search mechanism and interpretable adaptation of frequency-band quantization weights, improving exploration capability relative to discrete heuristic optimisation methods while maintaining full compatibility with the conventional JPEG decoding architecture. The resulting learned quantization matrices reflect dataset-specific frequency behaviour, supporting improved RD performance while preserving the structural principles of transform-based compression.

IV. EXPERIMENTAL SETUP

The proposed QWIO-JPEG framework is experimentally evaluated against the standard JPEG luminance quantization matrix under identical encoding conditions to assess its RD behaviour and perceptual reconstruction quality. All experiments were conducted on a workstation equipped with a single NVIDIA RTX 6000 Ada graphics processor with 48 GB of memory, an Intel Xeon W9 3595X processor, and 512 GB of RAM memory.

The optimization and reconstruction processes operate entirely in the DCT domain without altering the JPEG encoder or decoder, ensuring that all compressed images remain fully compatible with conventional JPEG-viewing systems.

To control computational complexity during optimization, images were first processed at a proxy resolution in which the longer dimension was resized while preserving structural content. The luminance channel of the proxy image was divided into 8 × 8 blocks and transformed into the frequency domain. Candidate quantization matrices generated through the proposed parameterization were evaluated using the QWIO optimizer, and after convergence the learned luminance quantization matrix was stored as $Q_{best}$. The same matrix was subsequently applied to the full-resolution image and reused for batch compression across the dataset.

The evaluation was performed across five visually diverse datasets to examine generalization across texture density, structural complexity, and frequency content. MNIST [21] contains grayscale handwritten digit images with simple geometric structure and sharp edges, making it suitable for analysing edge preservation under luminance quantization. CIFAR-10 [22] consists of small natural RGB images with moderate texture and color variation, providing a mid-level complexity benchmark. ImageNet [30] samples were used to assess performance on large-scale natural images with rich spatial frequency distribution. A brain-tumor medical imaging dataset [23] was included to evaluate compression in diagnostically sensitive regions with fine tissue boundaries, while the CASIA-Iris [24] dataset enabled assessment on near-infrared biometric images where circular edge structures dominate. Together, these datasets provide a spectrum ranging from low-frequency grayscale patterns to highly texture high-resolution natural images, enabling a robust evaluation of the optimized quantization table. Performance was measured using PSNR, SSIM, BPP, and pixel-wise error heat maps. Compression efficiency at the image level was computed using

$$BPP = \frac{S}{H \times W} \quad (16)$$

Where, $S$ denotes the estimated compressed size and $H \times W$ is the image resolution. Perceptual distortion localization was analysed through the absolute-error heat map

$$E(x, y) = |Y(x, y) - \hat{Y}(x, y)| \quad (17)$$

which highlights distortion concentration in edges, textures, and smooth luminance regions. Structural fidelity was quantified using SSIM, computed over local windows according to

$$SSIM(x, y) = \frac{(2\mu_x \mu_y + C_1)(2\sigma_{xy} + C_2)}{(\mu_x^2 + \mu_y^2 + C_1)(\sigma_x^2 + \sigma_y^2 + C_2)} \quad (18)$$

Where, $\mu_x$ and $\mu_y$ denote mean luminance values of the original image patch x and the reconstructed image patch y, respectively. $\sigma_x^2$ and $\sigma_y^2$ are the local variances of these patches, capturing contrast information, while $\sigma_{xy}$ denotes their covariance. The constants $C_1$ and $C_1$ are small positive stabilizing constants introduced to avoid numerical instability when the denominators approach zero; they are typically defined as $C_1 = (K_1 L)^2$ and $C_2 = (K_2 L)^2$, where L is the dynamic range of pixel intensities (e.g., 255 for 8-bit images) and $K_1, K_2$ are small scalar values commonly set to 0.01 and 0.03, respectively.

For every image across all datasets, two reconstructions were generated: one using the standard JPEG luminance quantization matrix and one using the optimized $Q_{best}$. In dataset-level evaluation, $Q_{best}$ was first learned from a representative subset of images and then reused to compress the remaining dataset, allowing assessment of its stability and reusability as a domain-adaptive quantization table rather than a per-image tuning artefact. This controlled and consistent evaluation protocol isolates the contribution of the optimized quantization matrix and provides a reliable basis for analysing the improvements produced by the quantum-walk-inspired optimization strategy.

## V. RESULTS AND DISCUSSION

This section presents a comprehensive performance of the proposed QWIO-JPEG framework on five datasets of ascending visual complexity: MNIST, CIFAR-10, ImageNet, Brain-Tumor medical images, and CASIA-Iris biometric images. Although the Quantum JPEG framework proposed in [19] introduces an innovative theoretical formulation for quantum-based JPEG compression, the study does not report quantitative experimental results or benchmark evaluations. Therefore, it is not feasible to carry out a comparison analysis on it to prove its efficiency relative to the proposed conceptual framework of QWIO-JPEG. Comparisons are drawn against the baseline JPEG scheme using the default luminance quantization matrix, with all other encoding parameters fixed. The performances are measured in terms of PSNR, SSIM, BPP, and qualitative visualization of error heatmaps. It is observed that the proposed method yields superior RD performance in all datasets, thus proving that the quantum walk inspired optimization learns quantization structures more consistent with frequency characteristics of each dataset. Table 1 presents the average PSNR comparison between baseline JPEG and QWIO-JPEG. The results indicate that the proposed approach achieves significant reconstruction quality improvements across all datasets.

*TABLE I : AVERAGE PSNR COMPARISON*

| Dataset | Baseline JPEG (dB) | QWIO-JPEG (dB) | Improvement |
|---|---|---|---|
| MNIST [21] | 27.60 | 31.20 | +3.6 dB |
| Cifar-10 [22] | 27.53 | 32.75 | +5.22 dB |
| ImageNet [30] | 25.08 | 35.45 | +10.37 dB |
| Brain-Tumor [23] | 41.62 | 46.79 | +5.17 dB |
| CASIA-Iris [24] | 44.36 | 48.79 | +4.43 dB |

The highest PSNR gains were observed on ImageNet and Cifar-10 datasets, where the quantum-inspired optimizer effectively preserved low and mid-frequency DCT coefficients while suppressing perceptually insignificant high-frequency components. Structural fidelity was further validated using SSIM, as summarized in Table 2.

The SSIM results confirm that QWIO-JPEG consistently improves perceptual similarity across all datasets, especially in those that are dominated by strong structural patterns like handwritten digits and iris textures. These improvements indicate superior preservation of contrast, luminance structure, and spatial coherence. Compression efficiency was measured using BPP, as shown in Table 3.

*TABLE II: SSIM COMPARISON TABLE*

| Dataset | Baseline JPEG | QWIO-JPEG |
|---|---|---|
| MNIST [21] | 0.9722 | 0.9859 |
| Cifar-10 [22] | 0.9697 | 0.9907 |
| ImageNet [30] | 0.8769 | 0.9876 |
| Brain-Tumor [23] | 0.9610 | 0.9846 |
| CASIA-Iris [24] | 0.9931 | 0.9978 |

The proposed QWIO-JPEG framework improves reconstruction quality significantly in terms of PSNR and SSIM while maintaining stable BPP values, thereby achieving an improved RD balance across diverse image domains.

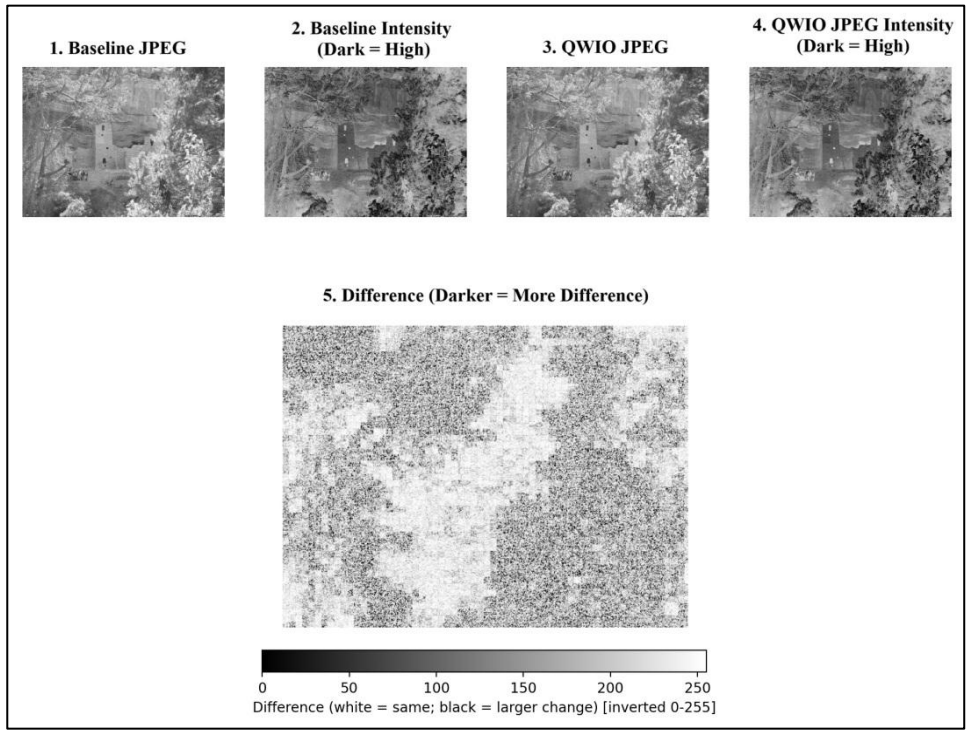

Figure 3 Figure illustrating sample ImageNet results, showing baseline JPEG and QWIO-JPEG compressed images, their intensity heat maps, and the corresponding error heat maps.

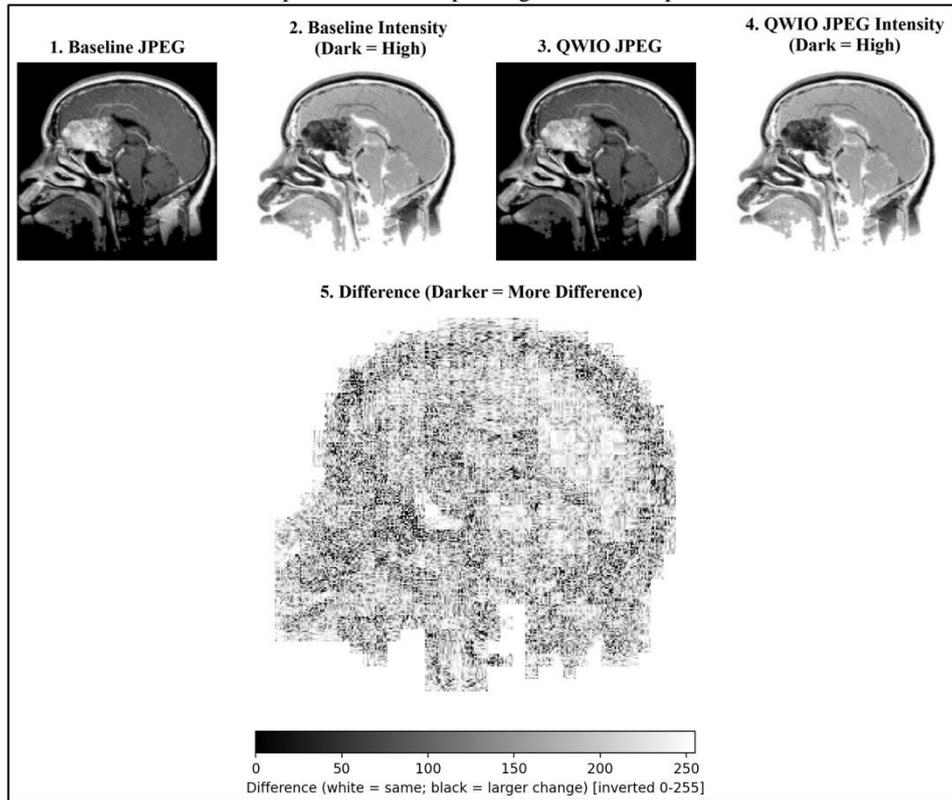

Figure 4: The illustrating sample Brain Tumor dataset results, showing baseline JPEG and QWIO-JPEG compressed images, their intensity heat maps, and the corresponding error heat maps.

*TABLE III: BITS-PER-PIXELS COMPARISON*

| Dataset | Baseline JPEG | QWIO-JPEG |
|---|---|---|
| MNIST [21] | 11.235 | 11.531 |
| Cifar-10 [22] | 10.641 | 10.751 |
| ImageNet [30] | 4.989 | 5.176 |
| Brain-Tumor [23] | 1.001 | 1.042 |
| CASIA-Iris [24] | 1.054 | 1.068 |

Results from qualitative error maps for pixel-wise errors correlate well to the results from the numeric measurements. On error maps of the baseline reconstruction of JPEG, errors tend to concentrate along object boundaries and textural regions, which suggests over-quantization of significant coefficients. Comparing the error maps of QWIO-JPEG to that of baseline reconstruction of JPEG, the error maps for QWIO-JPEG contain more balanced distributions of errors and smaller error values, perpendicular to distorted object boundaries and textural regions. CASIA-Iris images depict more balanced circular iris patterns and radial textural characteristics. Figure 3 illustrates an ImageNet instance: the figure shows the original and compressed images along with their intensity heat maps and the error heat map for the JPEG compressed and QWIO-JPEG compressed image. This map clearly reveals how QWIO-JPEG reduces details of edges and textures as compared to JPEG. In figure 4, an instance of an MRI of a brain tumor is illustrated for better visualization of the discussion, where the error is maintained at a lower value around the edges of the tumor, hence the structure is preserved.

The visual results indicate that quantum walk–inspired optimization redistributes quantization error across frequency bands in a perceptually meaningful manner, while the amplitude-driven search enables escape from local minima and identifies smooth, data-robust quantization patterns. Improvements in PSNR, SSIM, and BPP, together with reduced distortion in error heat maps, demonstrate the effectiveness of QWIO-JPEG in enhancing classical JPEG compression with full standard compatibility. Moreover, the reuse of a dataset-optimized quantization matrix highlights its potential as a reusable adjustment rather than an image-specific, one-time solution.

## VI. Conclusion & Future Research Directions

In this study, an adaptive quantum matrix for JPEG compression has been developed using a quantum-walk-based optimization function. A probabilistic, amplitude-driven search has been integrated with the standard JPEG framework to obtain a dataset-specific matrix that improves image reconstruction without compromising compression efficiency. Experiments have been conducted on five standard image datasets and have demonstrated improved reconstruction performance in terms of PSNR, SSIM, and per-pixel error heat maps. The proposed method has remained fully interoperable with standard JPEG decoders, enabling practical deployment. Future work has been planned to incorporate chrominance optimization, perceptual loss, real-time acceleration, and quantum–classical hybrid optimization.